\documentstyle[12pt]{article}
\title{Excitation energy of superdeformed bands in 
Relativistic Mean Field Theory}
\author{G.A.  Lalazissis$^{1,2}$ and P. Ring$^1$ \\
Physik Department, Technische Universit\"at M\"unchen \\
D-85747 Garching, Germany\\
$^2$ Department of Theoretical Physics\\
Aristotle University of Thessaloniki\\
GR-54006 Thessaloniki, Greece}

\begin{document}
\maketitle
\begin{abstract}
Constrained Relativistic Mean Field (RMF) calculations have
been carried out to estimate excitation energies
relative to the ground state for superdeformed bands in 
the mass regions A $\sim$ 190 and A $\sim$ 150.
It is shown that RMF theory is able to successfully reproduce 
the recently measured superdeformed minima in Hg and Pb nuclei.
\end{abstract}

Superdeformation has become the recent years one of the most interesting
topics of nuclear structure studies. After the well known area in the 
vicinity of the mass number A $\sim$ 150 a second region with 
A $\sim$ 190 has been discovered, where an impressive number of 
results has been obtained. 
However, despite the rather large amount of experimental information on 
superdeformed bands, there are still a number of very interesting 
properties, which have not yet been measured. A characteristic example
is the excitation energy between ground and superdeformed
bands. The excitation energy and the well depth of the superdeformed minimum
are amongst the most important factors which affect the decay of the
superdeformed bands to the ground state. 

Very recently discrete $\gamma$ rays directly connecting states of a
superdeformed band in $^{194}$Hg to the yrast states have been discovered
\cite{Kho.96}. 
This has made it possible to determine accurately the excitation energy
of the levels of the superdeformed band above the yrast line. 
Extrapolating to zero angular momentum the superdeformed 
minimum was found to be 6.017 MeV above the ground state. 
Similar measurements at about the same time
have also been reported for the superdeformed band of $^{194}$Pb nucleus 
\cite{Brinkman.96}. The excitation energy between the SD band and the
ground state was estimated to be 4.471(6) MeV. 
  
There are several non-relativistic  theoretical predictions for the 
excitation energy relative to the ground state.
Such as Hartree-Fock calculations with
density dependent Skyrme  \cite{Krieger.92} or Gogny interactions 
\cite{Delaroche.89} as well as calculations
using the Strutinsky method built on a Woods-Saxon average potential 
\cite{Satula.91}. In this letter we report and discuss the predictions of 
the RMF theory for the excitation energies relative to the ground state 
of the SD bands of $^{194}$Hg and $^{194}$Pb. In addition predictions 
for nuclei in the rare earth mass region are also provided.
  
Relativistic Mean Field (RMF) \cite{SW.86} theory has
recently gained considerable success in describing various
facets of nuclear structure properties. With a very limited
number of parameters, RMF theory is able to give a quantitative
description of ground state properties of spherical and
deformed nuclei \cite{GRT.90,Rin.96} at and away from the stability
line \cite{SLR.93,LS.95,LSR.96}. Moreover good agreement with
experimental data has been found recently for collective
excitations such as giant resonances \cite{VBR.94} and for
twin bands in rotating superdeformed nuclei \cite{KR.93,AKR.96}. 
Afansajev et al, have recently, and with great success carried out a 
systematic investigation of the entire A$\sim$140 to A$\sim$150 
mass region \cite{AKR.96a}.

The starting point of RMF theory is a standard Lagrangian
density \cite{GRT.90}
\begin{eqnarray}
{\cal L}&=&\bar\psi\left(\gamma(i\partial-g_\omega\omega
-g_\rho\vec\rho\vec\tau-eA)-m-g_\sigma\sigma
\right)\psi
\nonumber\\
&&+\frac{1}{2}(\partial\sigma)^2-U(\sigma )
-\frac{1}{4}\Omega_{\mu\nu}\Omega^{\mu\nu}
+\frac{1}{2}m^2_\omega\omega^2\nonumber\\
&&-\frac{1}{4}{\vec{\rm R}}_{\mu\nu}{\vec{\rm R}}^{\mu\nu}
+\frac{1}{2}m^2_\rho\vec\rho^{\,2}
-\frac{1}{4}{\rm F}_{\mu\nu}{\rm F}^{\mu\nu}
\end{eqnarray}
which contains nucleons $\psi$ with mass $m$,  $\sigma$-,
$\omega$-, $\rho$-mesons, the electromagnetic field and
non-linear self-interactions of the $\sigma$-field,
\begin{equation}
U(\sigma)~=~\frac{1}{2}m^2_\sigma\sigma^2+\frac{1}{3}g_2\sigma^3+
\frac{1}{4}g_3\sigma^4.
\end{equation}

In the present work we have carried out constrained RMF calculations
by imposing a quadratic constraint H$^{\prime}$ \cite{RS.80}, for the 
quadrupole moment using a mean field Hamiltonian H$_{RMF}$ and minimizing 
instead of  H$_{RMF}$ 
the expectation value of H$_{RMF}$ + H$^{\prime}$  where 
\begin{equation}
H^{\prime} = {c \over 2} ( <Q> - q)^{2}
\label{constraint}
\end{equation}
and q is the actual value of the quadrupole moment.
This constraint enables us to trace the energy surface  as
function of the quadrupole moment. 

For our calculations  we have adopted the frequently used parameter set 
NL1 \cite{Rei.89,RRM.86}, along with the recently proposed parametrization 
NL3 \cite{LKR.97} using a new version of the ``axially deformed'' code \cite{RGL.97}. The values of the two parameter sets 
are listed in Table 1. In the calculations  pairing has been included
within the BCS formalism. The experimental odd-even mass differences
were used to determine for the ground state gap parameters 
$\Delta_p$ (for protons) and $\Delta_n$ (for neutrons). These values
were then used to determine the corresponding monopole pairing constants 
$G_p$ and $G_n$ using the energy cut-off of $82 A^{-1/3}$ MeV. 
In order to treat properly the pairing correlations on going to the 
superdeformed minimum the corresponding gaps were not kept constant 
but calculated using these strength parameters 
for each value of the constraint.   

In Fig. 1 we show the potential energy surface for the nucleus
$^{194}$Hg as a function of the quadrupole parameter $q$ in Eq. 
(\ref{constraint}), calculated with the parameter sets NL3 and NL1. 
It is seen that for both Lagrangian parametrizations the RMF theory 
predicts an oblate shape for the the ground state of  the nucleus
$^{194}$Hg . This is  in accordance with experiment \cite{Drac.88}.
The calculations predict in addition a stationary point at
prolate deformation. In an axially symmetric calculation it looks
like a local minimum, but including triaxial deformations it 
corresponds probably to a saddle point connected to the oblate 
minimum by a valley in the $\beta$-$\gamma$ plane, which avoids the 
maximum at $\beta=0$.
For this nucleus the excitation energy $E_{x}$ of the superdeformed
minimum with respect to the oblate ground state is 5.62 MeV for the
parameter set NL1. It is  close to the measured value (6.02 MeV) 
\cite{Kho.96}. The parameter set NL3 does even better giving a value 
of 6 MeV in excellent agreement with the experiment. 
The depth of the potential well of the isomeric superdeformed state 
is about 2.66 MeV for NL1 and 1.25 MeV for NL3.

We have also performed similar calculations for the nucleus $^{192}$Hg. 
These are shown in Fig. 2 . For this nucleus the excitation energy 
$E_{x}$ is 3.93 MeV for NL1 while 4.46 MeV for NL3. The depths of the 
isomeric wells were found to be 2.70 MeV and 1.17 MeV respectively.

It is seen from Figs. 1 and 2 that going from $^{194}$Hg to $^{192}$Hg 
the excitation energy $E_{x}$ as well as the depth of the superdeformed
minimum are decreased for both parametrizations.
For NL1 the relative decrease is 30 \% and 1.5 \% respectively 
while for NL3 it is 26 \% and 6.5 \%. Moreover, it is also seen that 
for the two Hg nuclei NL3 predicts higher excitation energies and 
shallower potential wells as compared with NL1. An interesting feature 
which is common for both parametrizations is the fact that the product 
of the well depth and width of the superdeformed barrier
decreases by about 20\% on going from $^{192}$Hg to $^{194}$Hg.   
This quantity is relevant for the tunneling probability between
the superdeformed and the other minima and therefore for the
life time in the superdeformed well. Similar observations have
also been made in experimental investigations \cite{Jan.90}.
It should be noted at this point that the parameter set NL3 
gives values for the ground state binding energies which are in excellent 
agreement with the empirical values. Specifically NL3 predicts 
for $^{192}$Hg 1519.87 MeV and for $^{194}$Hg 1535.90 MeV the 
experimental values being 1519.20 MeV and 1535.50 MeV respectively. 
The corresponding predictions of NL1 are 1524.99 and 1540.31 MeV 
respectively. 

Non-relativistic Density Dependent Hartree-Fock calculations
predict values for the excitation energies $E_{x}$ of $^{194}$Hg,
which deviate somewhat from the experimental values. For Skyrme forces 
one finds \cite{Krieger.92} 5.0 MeV, while the Gogny's force
\cite{Delaroche.89} predicts 6.9 MeV. Finally, using a Woods-Saxon 
potential \cite{Satula.91} a value of 4.6 MeV was obtained. 
We can conclude that the predictions of RMF theory are in
better agreement with experiment than other theoretical 
studies. 

Our calculations stayed strictly in the mean field approximation.
There were no corrections of spurious rotational contributions
subtracted. Those corrections would influence in some way or
another the energy of all the stationary points in the energy
surface. However only the differences between the minima would 
be of importance. To calculate these contributions in an appropriate
way is a very difficult task, which definitely goes beyond the
present state of the art.
 
We next, carried out calculations for the nucleus $^{194}$Pb 
using the NL3 force. The potential energy landscape is shown in 
Fig. 3 (left side). RMF theory predicts for the excitation 
energy $E_{x}$ a value 4.53 MeV, which is very close to the 
experimentally measured value 4.471(6) MeV. The depth of the
isomeric well was found to be 2.2 MeV. The Hartree-Fock calculations with
a Skyrme interaction give an estimate of 4.86 MeV, which is also in good 
agreement with the measured value. The Woods-Saxon potential gives a lower
value of 3.8 MeV. In contrast to the $Hg$-nuclei we observe here
a rather sharp maximum between the ground state minimum at zero
deformation and the superdeformed minimum. However, this does not
mean pairing correlations vanish at this saddle point. As expected
we observe a local increase of pairing in going over the saddle
because of the increasing level density at this point.

Recently its was observed \cite{Gall.95} that the superdeformed band 
of $^{194}$Pb is populated at lower spin values than that of $^{192}$Hg 
and de-excites towards normally deformed states at a spin value lower than 
that of $^{192}$Hg.
This fact  suggests that the barrier separating the superdeformed and
normal deformed minima in the potential energy surface is higher in $^{194}$Pb 
than in $^{192}$Hg. This agrees with the predictions of the RMF theory.
Our calculations using the force NL3 for these two isotonic nuclei (N=90) 
show that the  height of the barrier of the nucleus $^{194}$Pb is about 1 MeV
higher than $^{192}$Hg and also the excitation energy $E_{x}$ of  $^{194}$Pb 
is slightly higher than that of $^{192}$Hg.

Superdeformed nuclei in the rare earth region are of great interest 
and there are projects for the measurement of the excitation energies 
$E_{x}$ \cite{Lieder.96}. We have therefore, also performed 
calculations for some even-even rare earth nuclei which exhibit 
superdeformed bands. In Table 2 we list the 
predictions of RMF theory for the excitation energy $E_{x}$. 
It is seen for Gd and Dy isotopes the $E_{x}$ values increase with  
increase of the mass number. However, the $E_{x}$ energies of 
$^{146}$Gd and $^{148}$Dy nuclei appear to be larger than their 
neighboring counterparts. This could attributed to the magic character
of these nuclei which are expected to have a deeper minimum.  
Moreover, it is observed that the isotonic nuclei, $^{142}$Sm, $^{144}$Gd,
(N=80) , $^{148}$Gd, $^{150}$Dy (N=84), $^{150}$Gd, $^{152}$Dy (N=86) have
rather similar excitation energies $E_{x}$. 

In the same table we show predictions of RMF theory for the ground 
state (g.s) binding energy (BE) together with the corresponding
empirical values (BE$_{exp.}$), taken from the most recent compilation of
 Audi and Wapstra \cite{Audi.96}. It is seen that RMF theory with the 
effective force NL3  is able to reproduce  the experimental values with 
very high accuracy, the disagreement being less than 0.1 \%.
The excellent predictions for the  g.s binding energies give us confidence
for the correctness of our estimate for the excitation energy $E_{x}$.
Finally, we show in Fig. 3 (right side) as an illustration the energy curve of 
the nucleus $^{146}$Gd calculated with the parameter set NL3. 
The excitation energy $E_{x}$ is 10.63 MeV and the depth of the
superdeformed minimum is 0.66 MeV.

In summary, we have performed constrained RMF calculations using 
the effective forces NL1 and NL3 for several nuclei in the 
A $\sim$ 190 and A $\sim$ 150 regions to estimate the excitation 
energy of the superdeformed minimum relative to the ground state. 
It has been shown that RMF theory is able to reproduce the recently 
reported values for the excitation energies of $^{194}$Hg and 
$^{194}$Pb nuclei with a higher precision than other non-relativistic
calculations. Moreover predictions for the heights of the superdeformed 
barrier in the A $\sim$ 190 mass region have also been given. The 
calculated values agree with scenarios suggested by experimental 
finding. Finally, RMF theory predictions have been also made for the 
excitation energy $E_{x}$ of the well known superdeformed nuclei 
in the rare earth mass region, which should also be useful for comparisons 
with future experiments.

\bigskip
This work has been supported by the Bundesministerium f\"ur 
Bildung und Forschung under the project 06 TM 785. We thank
R. M. Lieder for useful discussions.

\newpage
%%%%%%%%%%%%%%%%%%%%%%%TABLE 1%%%%%%%%%%%%%%%%%%%%%%%%%%%%%%
%%%%%%%%%%%%%%%%%%%%%%%%%%%%%%%%%%%%%%%%%%%%%%%%%%%%%%%%%%%

\begin{table}
\caption{The parameter sets of the effective interactions NL1 and NL3 of 
the Lagrangian of the RMF theory
used in the present work. The masses are given in (MeV) and the coupling 
constant $g_2$ in (fm$^{-1}$) }
\begin{center}
\begin{tabular}{llllllllll}
\hline
  & $M$ & $m_{\sigma}$& $m_{\omega}$& $m_{\rho}$ & $g_{\sigma}$ &$g_{\omega}$
  & $g_{\rho}$&   $g_2$ &  $g_3$ \\
\hline
\      \\
NL1 & 938 &  492.250  & 795.359 & 763.0& 10.138 & 13.285 & 4.976& -12.172&
-36.265 \\
\        \\   
NL3 & 939 &508.194 &  782.501 &  763.0&10.217  &  12.868&  4.474  & -10.431
 &  -28.885 \\  
\hline
\end{tabular}
\end{center}
\end{table}
%%%%%%%%%%%%%%%%%%%%%%%%%%%%%%%%%%%%%%%%%%%%%%%%%%%%%
%%%%%%%%%%%%%%%%%%%TABLE 2%%%%%%%%%%%%%%%%%%%%%%%%%%%
%%%%%%%%%%%%%%%%%%%%%%%%%%%%%%%%%%%%%%%%%%%%%%%%%%%%%
\begin{table}
\caption{The predictions of the RMF theory for the excitation energy $E_{x}$
relative to the ground state of some Sm,  
Gd and Dy isotopes using the effective force NL3. Also shown are the calculated
and empirical values of the (g.s) binding energies.}
\begin{center}
\begin{tabular}{llll}
\hline
 Nucleus &BE & BE$_{exp}$&  $E_{x}$ \\
\hline
 $^{142}$Sm &1177.02 &1176.62 & 6.80 \\
 $^{144}$Gd &1184.62 &1184.12 & 6.45 \\
 $^{146}$Gd &1205.43 &1204.44 &10.63 \\
 $^{148}$Gd &1220.91 &1220.77 & 7.82 \\
 $^{150}$Gd &1236.85 &1236.40 & 8.00 \\
 $^{148}$Dy &1211.47 &1210.75 &10.32 \\
 $^{150}$Dy &1228.35 &1228.39 & 7.98 \\
 $^{152}$Dy &1245.69 &1245.33 & 8.32 \\
 $^{154}$Dy &1262.94 &1261.75 & 9.90 \\
\hline
\end{tabular}
\end{center}
\end{table}
%%%%%%%%%%%%%%%%%%%%%%%%%%%%%%%%%%%%%%%%%%%%%%%%%
%%%%%%%%%%%%%%%%%%%%%%%%%%%%%%%%%%%%%%%%%%%%%%%%%
%%%%%%%%%%%%%%%%%%%%%%%%%%%%%%%%%%%%%%%%%%%%%%%%%
%
%           Figure Captions
%
%%%%%%%%%%%%%%%%%%%%%%%%%%%%%%%%%%%%%%%%%%%%%%%%%
\leftline{\Large {\bf Figure Captions}}
\parindent = 2 true cm
\begin{description}

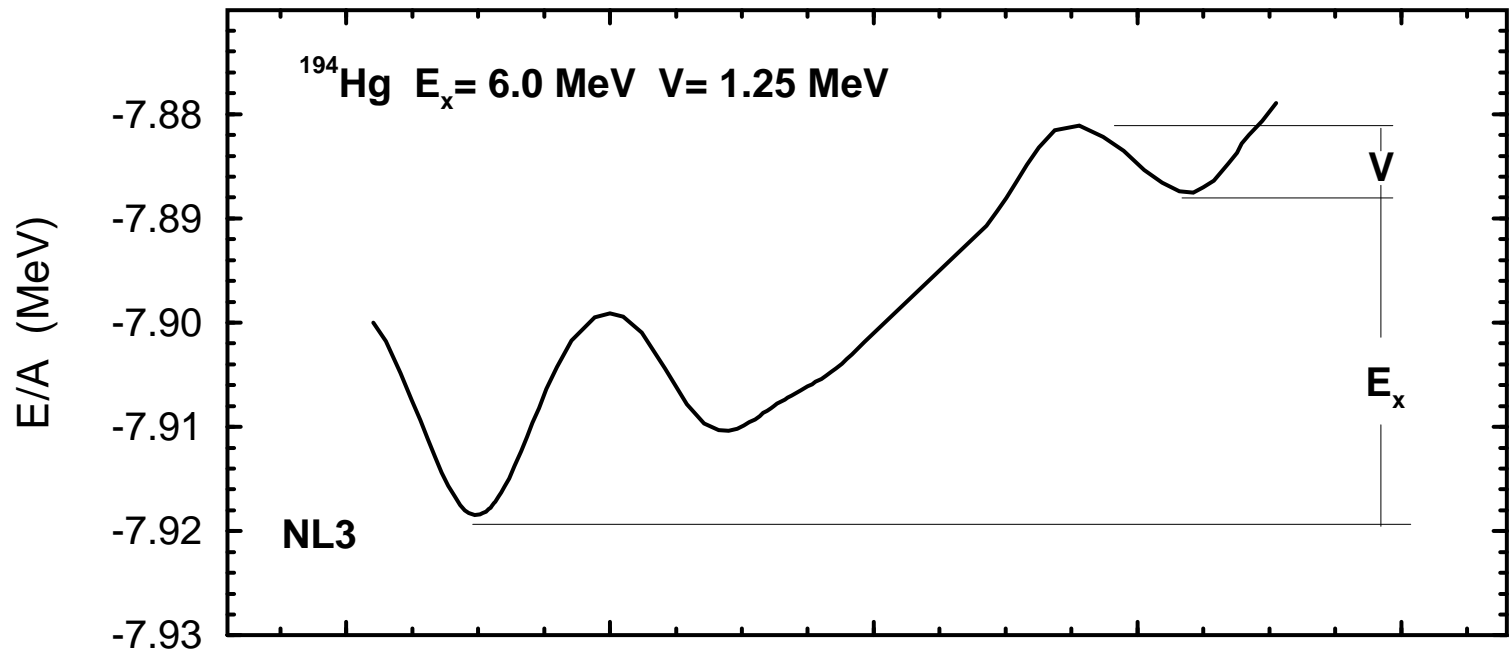
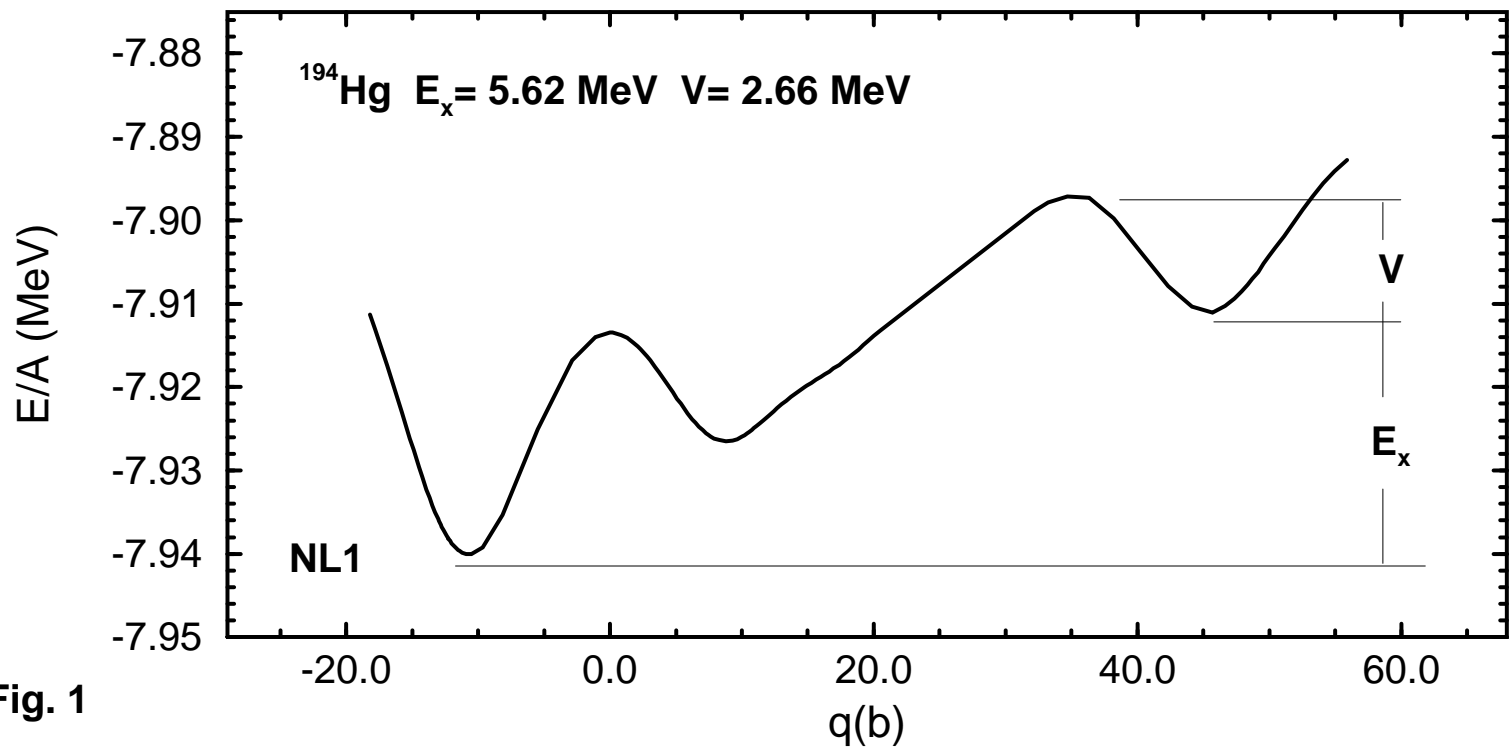
\item[Fig. 1] Potential energy surface as a function of
the quadrupole parameter $q$ for the nucleus $^{194}$Hg
calculated using two different parameter sets NL3 (upper part)
and NL1 (lower part).

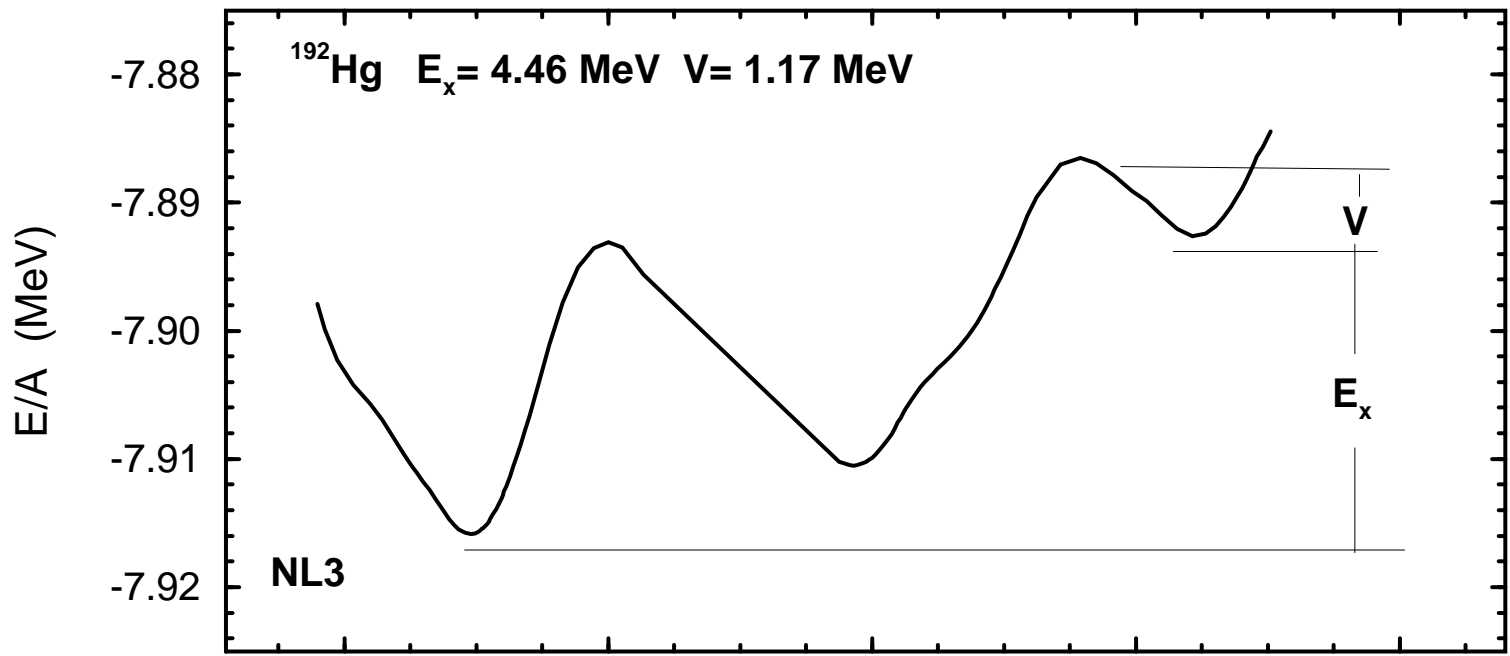
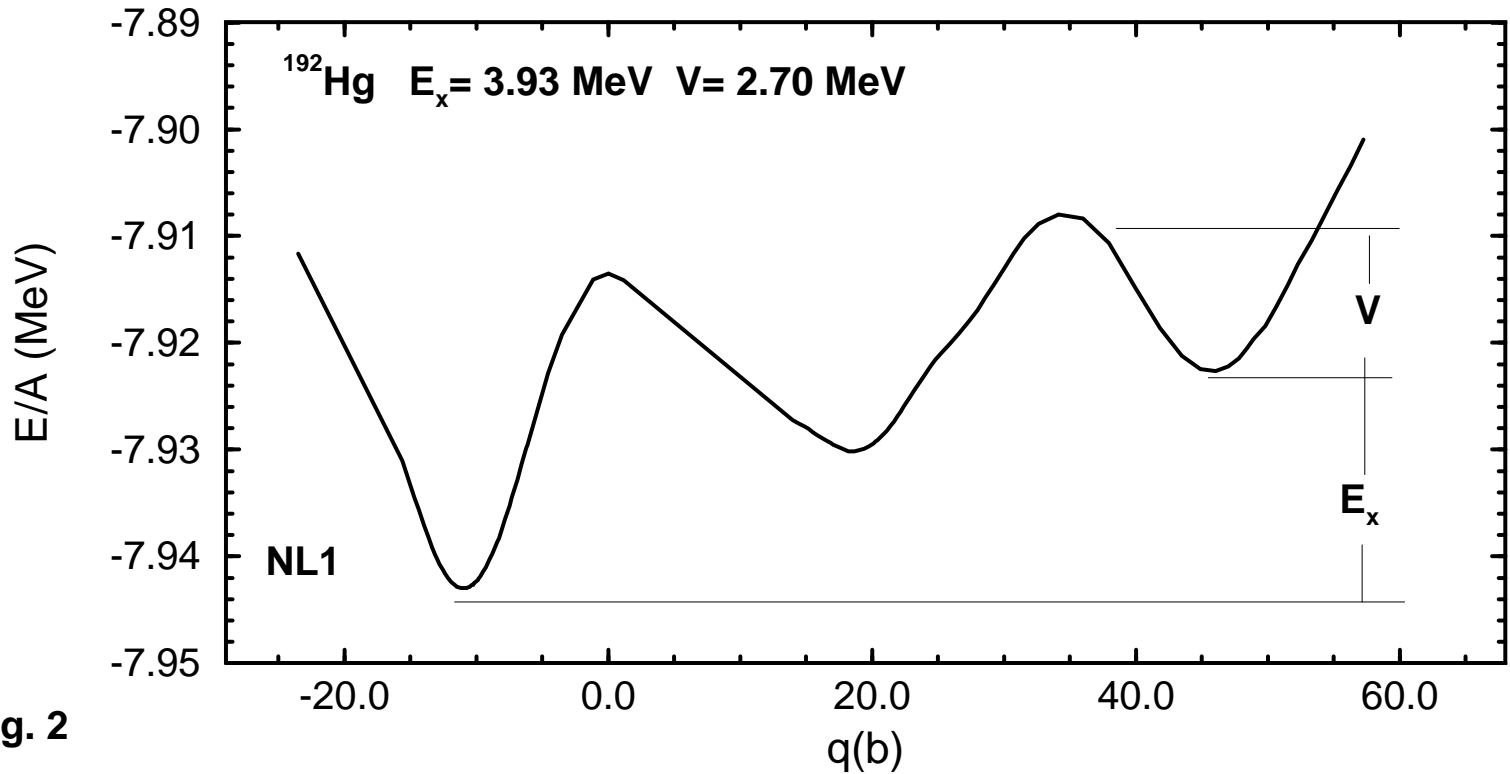
\item[Fig. 2] Potential energy surface as a function of
the quadrupole parameter $q$ for the nucleus $^{192}$Hg
calculated with two different parameter sets NL3 (upper part)
and NL1 (lower part).

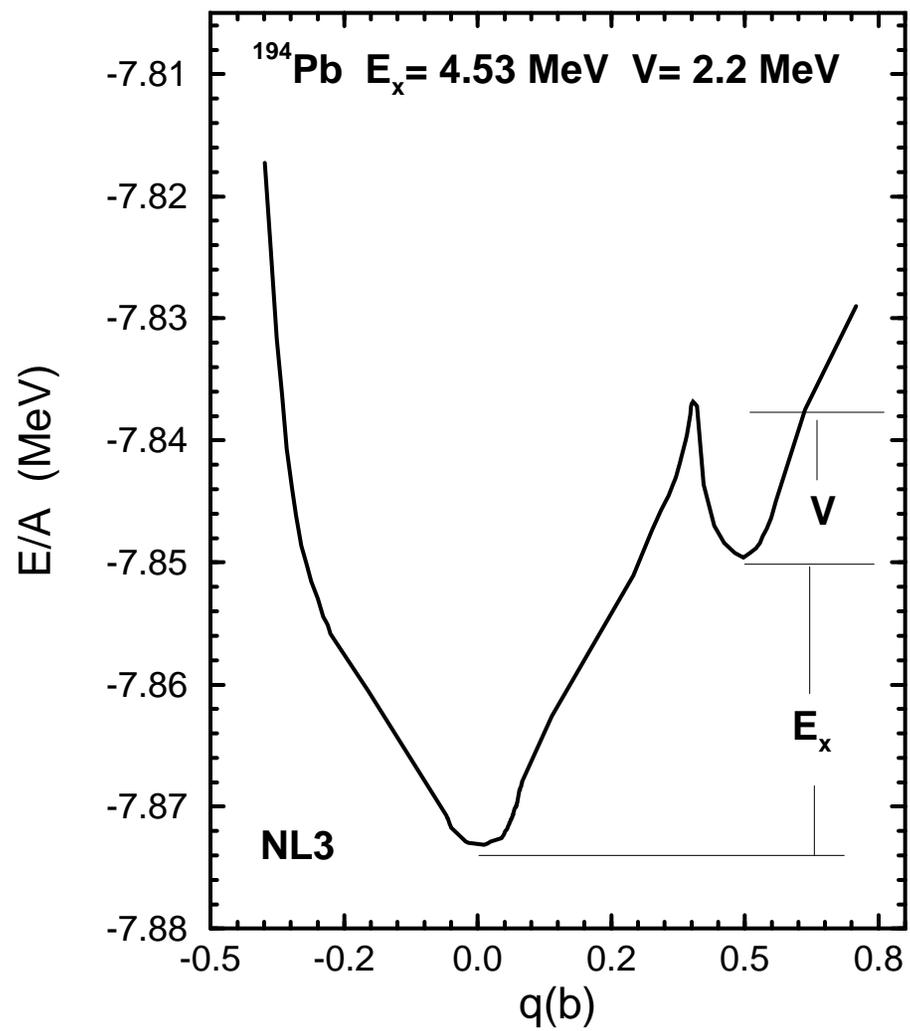
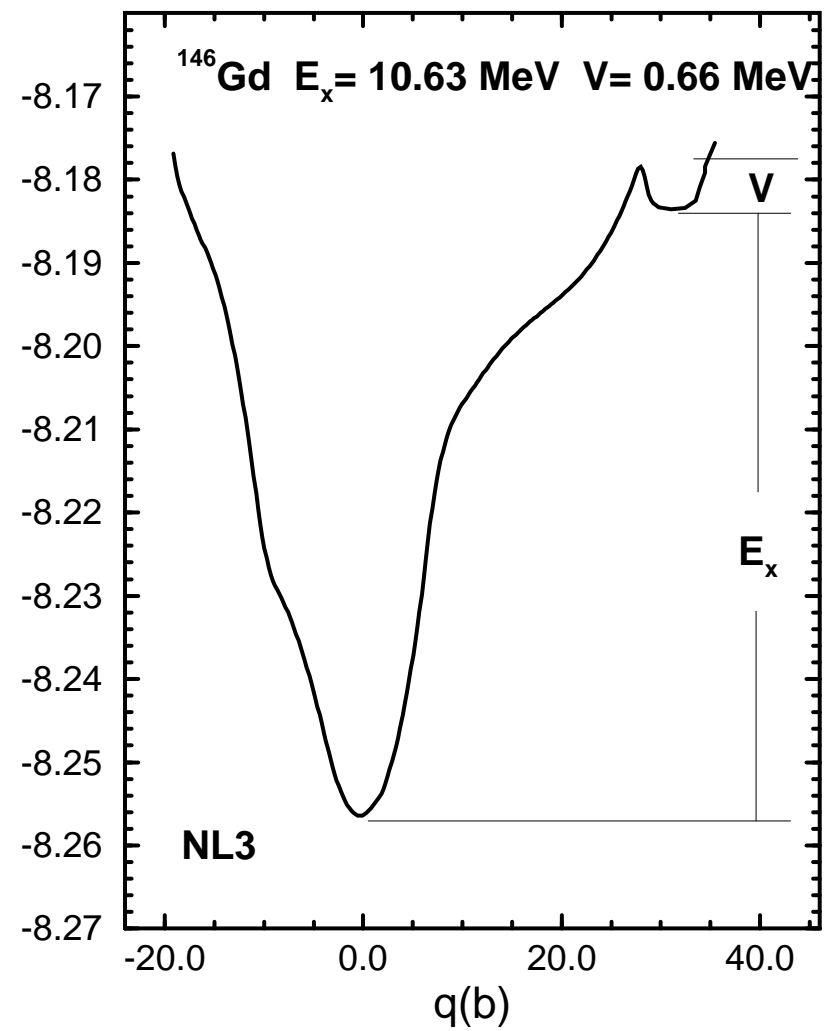
\item[Fig. 3] Potential energy surface as a function of
the quadrupole parameter $q$ for the nuclei $^{194}$Pb (l.h.s)
and $^{146}$Gd (r.h.s.)
\end{description}

\end{document}